# Protein Sequence, Structure, Stability and Functionality

J. C. Phillips

Dept. of Physics and Astronomy, Rutgers University, Piscataway, N. J., 08854-8019

## Abstract

Protein-protein interactions (protein functionalities) are mediated by water, which compacts individual proteins and promotes close and temporarily stable large-area protein-protein interfaces. Proteins are peptide chains decorated by amino acids, and protein scientists have long described protein-water interactions in terms of qualitative amino acid hydrophobicity scales. Here we examine several recent scales and argue plausibly (in terms of self-organized criticality) that one of them should be regarded as an absolute scale (within the protein universe), analogous to the dielectric scale of bond ionicity in inorganic octet compounds. Applications to repeat proteins (containing upwards of 900 amino acids) are successful, far beyond reasonable expectations, in all cases studied so far. While some of the results are obvious and can be obtained from the *ex vitro* spatial structures alone, many are hidden from plain view, and can be called phantom relations. As a byproduct, the network theory explains the exceptional functionality of leucine in zippers, heptads, and repeat consensus sites.

**1 Introduction**

While the centrality of structure-function relationships to our understanding of nature has been accepted since Plato's time (he proposed that the structures of the four known elements [earth, air, fire and water] were those of the four smallest Archimedian polyhedra), it was not until 1856-7 that Kekule introduced the (then controversial) concept of tetravalent carbon, thereby founding modern organic chemistry. He followed this in 1865 with the idea of double valence bonds in benzene rings, establishing the basis for structure-function relationships in aromatic hydrocarbons. However, this success soon led to contradictions in trends predicted by the planar polycyclic structures alone (some carbons have two neighbors, some have three). The contradictions were resolved when it was realized (again after considerable controversy) that the properties of aromatic hydrocarbons were not determined primarily by their strongest planar σ bonds, but rather by their much weaker π bonds. (Hueckel's π electron quantum theory, developed by him in 1930-31, was not generally appreciated until it was rediscovered by others in 1950-51! This point – that physical properties are often determined by *weak*, not strong, interactions, is widely overlooked even today, for example, not only in proteins, but also in "strong interaction" models of the electronic properties of inorganic solids, such as superconductivity.) Finally, Pauling laid the foundations of molecular biology in 1951 with his discovery of the α helix[1].

Structure-function relationships in proteins are complex for both fundamental and practical reasons. One should first distinguish between stability (ground state) and functionality (response functions). These two quite different quantities can easily exhibit qualitatively different chemical trends, which are important for many purposes, for example, in quantifying homology analyses. These contrary trends can also arise from fully deductive ("first principles") quantum methods that optimize the ground state energy at the expense of excited state energies and properties. Wigner and Slater first appreciated contradictions of this kind in the 1930's, but they were not widely recognized until some 50 years later. Today several complex techniques are known for dealing with this problem, but they are limited to few atoms[2].

The prototypical example of a combined method is the theory of ionicity of chemical bonds; here a clever combination of deductive and inductive methods produced extremely accurate results, which have not been replicated by deductive or inductive methods alone. The discussion is restricted to the context of simple binary octet compounds $A^N B^{8-N}$, where there are two broad classes to be considered[3,4]. In six-fold coordinated salts with ionic bonding (NaCl, CaO), increasing binding energy reduces polarizability, but in four-fold coordinated semiconductors with covalent bonding (Si, GaAs), increasing binding energy reduces the average energy gap and increases polarizability (due to quantum interference of atomic orbitals in the bonding overlap region)[2,3]. By exploiting the deductive f-sum rule, the dielectric theory of ionicity derives a *dimensionless* fractional ionicity that is *100% successful* in predicting the structural separation of these compounds, using as inductive input only the valence electron density and the electronic polarizability. The theory also explains many other properties of these (technologically important, hence well-studied) materials, including the collapse of shear resistance (called enhanced conformational flexibility by protein scientists) in the AgX salts (X = Cl, Br, I) as the ionicity increases to its critical value[3,4].

The essentially complete success of the dielectric theory of ionicity is based on several subtle points. Had the *energies* in the ionic salt and covalent semiconductor structures been calculated separately, the large cancellations in most of the energy differences would have amplified the effects of small errors in either first principles quantum calculations, or (worse) parameterized classical force field (CFF) models, leading to ~70% accuracy at best. (These problems were already evident in CFF (Born-Mayer) models of binary salts in the 1930's, which is why they were abandoned in favor of quantum theories[5].) The use of the electronic polarizabilty to construct covalent and ionic energy gaps turned out to be especially felicitous, as the latter is measured at optical wavelengths, and weights long-range *weak* interactions more heavily compared to short-range nearest neighbor strong ones.

Today the successes of the dielectric scale of ionicity are so extensive that it is regarded as the Kelvin scale (analogous to his absolute temperature scale) of quantum chemical calculations, which presently are able to reproduce its results with 90% accuracy[6]. Here we argue that a properly constructed scale of hydrophobicity exists that is also *dimensionless* and can be expected to be almost equally successful on the much larger protein-water canvas. Of course, the success of such a scale must be confirmed in practice, but the protein-water canvas contains hundreds of thousands of proteins, and is so large that even hundreds of successful applications could be unconvincing. Here we analyze the fundamental features of proteins which combine to make such a scale possible in principle. As we shall see, the distinction between stability and functionality is crucial. Its place in the logical structure of protein science, especially protein theory, is indicated in Fig. 1.

## 2. Water – Protein Interactions are Critical to Protein Functionality

The success of a *dimensionless* ionicity scale for inorganic octet binary compounds is unlikely to convince protein scientists that a similarly *dimensionless* hydrophobicity scale can plausibly exist for proteins, which are much more complex. The first step towards establishing plausibility is the observation that proteins have evolved *in vitro* to form a contextual universe of their own. Thus the proposed amino acid hydrophobicity scale is restricted to the context of wild proteins, and to only weakly mutated proteins; it does not apply to isolated amino acids, and the *dimensionless* hydrophobicity scale should not predict transference *energies* of amino acids from water to an organic solvent such as octanol. (The latter are used to calibrate short-range hydrophobic interactions in CFF models, so it is already clear at this stage that the hybrid *dimensionless* hydrophobicity scale will address problems of functionality in quite a different way from CFF simulations.)

The second plausibility point is that protein functionality depends essentially on large-scale network reversibility. Dynamical processes in large-scale networks (such as polymers) are generally irreversible because of entanglement. A closely related question is the uniqueness of folding, sometimes described as Levinthal's paradox: how can a protein of complexity of order $10^N$ (with N ~ 300) either fold uniquely, or evolve from individual amino acids? The answer to this is that in the presence of alternating rigid α helices and floppy connecting loops, the dimension of the available configuration space greatly shrinks, from order $10^N$ to order $N10^C$, where C is the number of conformationally active degrees of freedom utilized in protein functionality. This general consideration is strongly supported by the observation that helical correlation lengths (for instance, in the repeat proteins discussed below, or indeed for any protein interacting with proteins other than itself) are generally of order 35 residues. Previously we noted that purely mathematical studies have shown that Brownian diffusion in N-dimensional spaces exhibits a critical instability near N ~ 40, and that "the overall sign [hydrophobic compaction vs. hydrophilic expansion] of a protein can he defined in terms of a product of curvature and hydrophobic(philic) character over all amino acid residues"; thus a stable protein usually can be built only out of secondary elements no larger than N ~ 40 residues[7].

There is strong support for this picture of reversibility in proteins from the recent discovery[8,9] of a narrow reversibility window in inorganic oxide and chalcogenide network glasses (such as $SiO_2$ and window glass). The window has abrupt edges, and glasses within the window exhibit such remarkable properties that these materials are said to form an Intermediate Phase (IP). One of the remarkable properties is that the usually nearly constant non-reversible enthalpy of the glass transition is nearly zero in the window, where aging also nearly ceases[8]. Another remarkable property is that the non-zero stress threshold seen in polymers, and described by the classic worm-like-chain model in terms of a free-volume model with excess configurational entropy, is present outside the window, but abruptly vanishes within it[9]. This same non-zero stress threshold has been observed by Atomic Force Microscopy[10] in *ex vitro* (denatured) DNA and in repeat proteins (discussed below), but it would presumably be nearly absent for collapsed proteins *in vivo*. The importance of the reversibility window, and its relation to the distinction between stability and functionality, is sketched in Fig. 2.

There is an important connection between the narrowness of the network glass reversibility window and the diffusion-limited stability of helical repeat lengths. If we think of protein dynamics as analogous to self-organized criticality[11], then such scale-free models immediately encounter the problem that helical lengths are *not* arbitrarily large. Instead of arbitrarily large lengths associated with a true critical point, what we have instead is a truncated spinodal (spindle-top), and associated with that truncation is a cut-off length. That cut-off is quite abrupt for most proteins, and the reversibility windows of network glasses (especially designed network glasses, like Ge-S-I) can be as narrow as 1% in alloy composition[12]; thus the spatial cutoff need not lead to irreversibility in network dynamics. It has been suggested that in network glasses such cutoff lengths may be the result of percolation of some kind of Zallen cluster[13,14] (analogous to α helices). In any case, as large scale (nearly critical) conformational motions dominate inter- and intra-cellular protein dynamics, it is already evident that it is the *weak*, long-range forces that are likely to be most important for protein functionality, and that these weak forces probably involve marginal hydro interactions.

Before we leave this subject, what about the specific geometrical problems involving protein α helical twists, and the fact that helical amide NH residues are hydrogen bonded to O=C at *i* and *i* + 4 (or 3 in a more compact 3-10 helix)? Such short-range packing motifs (for example, leucine zippers) are always identified in *ex vitro* crystalline structures, and even there their contribution appears to be most important for stability (not functionality)[15]. We will see examples later that indicate that some common functional effects associated with solvent exposure geometries, even of short turns, are easily quantified with accurate hydrophobicity scales. However, in keeping with the principle of treating the simplest cases first, the present discussion focuses on examples where hydrophobic packing effects are usually small, and the marginally attractive inter-helical interactions are water mediated.

## 3. Collective H Bond Interactions: Resonating H Bond Networks

The obvious analogues of the electronic polarizability that was used to construct the dielectric scale of ionicity for octet compounds are the infrared wave lengths and oscillator strengths of H bonds. Unfortunately, most of these amino-acid-specific vibrational spectra are not separable from the infrared spectra of other amino acids in proteins. Quantum theorists have fully appreciated the importance of these excitations as indicators of collective H bond interactions, and hundreds of studies[16] are available for small molecules that show large wave length shifts (both blue and red, -philic and -phobic) and especially gigantic oscillator strength enhancements that are inexplicable with CFF models.

Of course, quantum theory gives an accurate account of chemical trends in the stabilities of small hydrogen-bonded amides[17], including changes in amide H bond binding energies by as much as a factor of 10. The effects of collective H bond interactions on energy shifts and oscillator strengths are equally striking[18], with enhancements in infrared oscillator strengths of as much as a factor of 2000 in a chain of nine 4-pyridones compared to the monomer; a reasonable upper limit for a classical dipole model would be a factor of 100 or less. The importance of non-classical hydrogen-bond cooperativity in α helical stabilization energies and infrared oscillator strengths is an old story to quantum theorists[19] that is constantly being strengthened[20,21].

It should be emphasized that the fact that H bonding can cause both red- and blue-shifts in amide infrared vibrational modes is very similar to the fact that in octet compounds increasing binding energy can either increase or decrease polarizability[3,4]; it is direct evidence that H-bonded organic networks contain a delicate balance of amide ionic and covalent interactions. This delicate balance is analogous to the nearly complete softening of shear modes in octet compounds near the covalent-ionic transition. It explains the results of quantum calculations that show that the strength of amide H bonds is very sensitive to the internal conformation of polypeptide chains, where it was also noted that "there has been some thought that the energetics of H-bond formation of a polypeptide might be directly correlated with the dipole moment of a segment of the latter. However, an analysis of the computed data showed no evidence of any such correlation."[22] Not only can such interactions not be represented by simple electrostatic models, as demonstrated by the failure (1930's) of CFF models for the octet compounds, they cannot be represented accurately by any CFF.

It is interesting that, contrary to the results of all first-principles calculations, many molecular biologists continue to favor classical models of hydrogen bond cooperativity and amino acid solvation[23], if only because it is seen as the only practical approach to protein stability and the energetics of protein folding. (They also discuss the energetics of hydrophobicity both in terms of solvent transfer models and a "packing-desolvation" cavity model (similar to the Lorenz (~ 1900) models for internal electric fields), neither of which is consistent with our present approach to weak interactions, collective H bonding, criticality, and hydrophobicity scales.) Note, however, that in the octet valence compounds the dielectric ionicity scale predicts structures and cohesive energies far more accurately, not only than any (Born-Mayer) CFF, but also even more accurately than the best available first principles quantum calculations. Thus one could equally well reason that necessity impels us to search even more carefully for a similarly accurate, and equally sound, contextual protein amino acid hydrophobicity scale.

**4. Hydrophobicity Scales**

By now the reader should have been convinced of the desirability, and perhaps even the necessity, of finding accurate hydrophobicity scales, but it is clear that many such attempts have already been made, with disappointing results[23]. All of these failures can be traced to treating protein amino acid trends in terms of individual amino acid interaction energies, a reasonable place to start, and proposed historically by Kauzmann[23]. In the protein context (or "universe"[11]) one is in the vicinity of a critical point, and other approaches, based on expansions in terms of individual aa energies, aa pair energies, aa triplet energies, …, are (in effect) virial expansions about the ideal gas (zero aa density) limit; such expansions must fail near the covalent-ionic (or hydrophilic-hydrophobic, or collapsing – expanding)) transition relevant to protein functionality.

Two groups have appreciated the necessity for developing hydrophobicity scales entirely in a water-only, collective (implicitly self-organized) protein context. Pintar *et al.* have used the phenomenon of hydrophobic collapse itself to define multiple hydrophobicity scales S based separately on buried depths of aa's at helical sites and all sites[24]. This method does not utilize organic solvents, but it does assume that the protein structure determined in the "dry" crystal has not changed much from its functional structure in water. This is a reasonable assumption, as the optimized (densest) packing of the large secondary units is determined mainly by steric hindrance. Two hydrophobicity scales thus defined are shown for the aa "alphabet" [H = Glycine = G, $CH_3$ = Alanine = A, $(CH_3)_2CH$ = Valine = V, etc] in Fig. 3.

Pintar himself has expressed some reservations concerning his results: there are minor geometrical ambiguities associated with the fuzziness of the surface, and both his and later data show evidence for a crossover around 175 residues from surface-dominated to volume-dominated behavior[24,25]. These ambiguities arise because the buried scale has the dimensions of length $L$. Another issue is how the scale should be applied. It is simple enough to permit the construction of a helix-loop cluster dictionary based on five- (not one-or two-!) aa clusters[26]. This approach, however, does not distinguish between stability and functionality, and also mixing the cluster (virial) approach with criticality in the vicinity of a covalent-ionic transition is questionable.

The second approach is based on the idea that one can be more precise by cleverly utilizing the concept of solvent accessible surface areas (SASA)[27]. By itself this is an old idea: one determines the SASA of each aa by a Voronoi tessellation (polyhedral construction) based on Van der Waals radii, and then fixes the surface area accessible to a 1.4 A spherical probe (water molecule). This approach (dimension: $L^2$) alone is, however, inaccurate[23]. Now comes a highly original idea: the scale is based not on the average SASA seen by a given aa (**S**(1)), but rather on the way this SASA contracts with increasing N in a helical environment of 2N +1 aa's on a chain, $1 \leq N \leq 17$. (This corresponds to a virtual folding of each helix centered on a given aa, and it assumes that some kind of systematic behavior associated with chain hydrocurvature will emerge upon averaging overlapping SASA's over a large number of cases, here 5526 ultra-high resolution helical fragments.) In fact, each SASA contracts self-similarly, according to

$$SASA(aa) = Const(2N+1)^{-\gamma(aa)} \quad (1)$$

One can use $-\gamma(aa)$ as an "ideal" hydrophobicity scale: note that it does not have dimensions of either energy or length, in fact, like fractional iconicity, it is *dimensionless*; the corresponding scale is also shown in Fig. 3.

From the present point of view (proximity to a critical point, or more accurately, a spindle-top of a collective water-protein covalent-ionic spinodal), the self-similar behavior discovered by Zebende is not surprising, in fact, it is *just what we were looking for* in order to assure reversibility and minimal configurational entropy. Does it have a classical analogue? Yes, there is an incomplete one. The packings of amino acids in folded proteins are locally icosahedral and resemble those found very generally in granular materials, foams, froths, and glasses[28]. Numerical simulations[29] show that the growth of soap froth cells ("dynamical percolation") in Voronoi tessellations follow fractal power laws similar to those found above for SASA's, with the exponents reversing sign depending on the ordering of the initial centers. Presumably this model could be generalized to one with multi-colored starting points, and a different growth rate for each color. If the cells are allowed to interpenetrate, then those that grow faster will overlap those that grow more slowly, leading to fractal growth with different exponents. Growth could be arrested when the overlap reached a cutoff value. The main difference between this model and actual protein aa dynamics is that in the latter the aa's are constrained to form a chain. Whether or not this one-dimensional constraint (effectively combining the travelling salesman problem with soap film dynamics) would invalidate the fractal behavior is not easily decided (even by simulations), but for our present purposes Zebende's exponents look very convincing, and they are not likely to be improved upon soon.

Looking at Fig. 3, one's first reaction may be to say that there is considerable scatter between the second line (mean buried) values and the third-line (self-similar) values $-\gamma(aa)$. However, this scatter (listed as standard deviations in the figure caption) may simply reflect the differences between a dimensionless self-similar scale and a scale linear in $L$. The differences would presumably be even larger for the individual SASA (dimension: $L^2$). Also it is important to keep in mind that even the self-similar values $-\gamma(aa)$ should not be used in isolation, but always averaged in some way over groups of at least three aa's. One of the obvious internal correlations in aa sequences is the tendency towards hydroneutrality, and this makes the differences between such averages for the mean buried and self-similar scales usually (but not always) much smaller than would be expected if the individual differences fluctuated randomly.

Some time ago ("prehistoric" in a field that considers results from 10 years ago to be "classical"), a hydrophobicity scale based on "an amalgam of experimental observations [on isolated amino acids] derived from the literature" was developed[30]. It showed that the interior portions of the sequences of water-soluble globular proteins are largely hydrophobic, while the exterior portions are hydrophilic. In membrane-bound proteins, the portions of their sequences that are located within the lipid bilayer were also clearly delineated by large uninterrupted areas on the hydrophobic side of the midpoint line. Comparison of this scale with the mean buried and self-similar scales shows comparable mean-square differences, but the latter are noticeably superior in many small, but important, details; for example, the most hydrophilic aa's should be lysine (K) and arginine (R) (in that order, as in the buried and self-similar scales, but not the Kyte-Doolittle (KD) scale), and leucine should fall in the middle of the hydrophobic range (on the KD scale it is the most hydrophobic).

One of the striking statistical aspects of the 20 protein amino acids is that they occur with different frequencies. For packing reasons the smallest amino acids might be expected to occur with the highest frequency: these are glycine (G) and alanine (A), which are both hydroneutral on the Zebende scale. (They are both hydrophilic on the mean buried scale, and alanine is strongly hydrophobic on the KD scale.) Indeed G and A are the second- and third- most frequent protein aa's (~ 7-8%), but the most frequent aa is leucine (9.1%). To what does leucine owe this signal honor?

That is not all. Many repeats are dominated by leucine zippers or are intermediate leucine-rich repeats. Leucine heptad zippers represent a dense solenoidal helical "knobs into holes" packing motif that was identified almost immediately by Crick for dimers[31]. It has turned out to have many variants, all the way up to septamers, with a wide variety of packing combinations[32]. If Van der Waals packing interactions were the dominant factor in determining packing sequences, such a wide variety would seem unlikely, even in the absence of a reliable CFF. Instead, the observed diversity of packing themes, together with a consistent local alternation of hydrophobic and hydrophilic residues, suggests that these closely packed structures are still influenced by short-range water-mediated interactions. However, from the viewpoint of theory, treatment of such short-range interactions requires not only a long-range hydrophobicity scale, but also amino-acid-specific shorter range scales, an issue that lies beyond the scope of this work.

Leucine-rich repeats (LRR) are an intermediate case, mixing thick α and thin 3-10 helices in a curved cylindrical packing geometry. Quite generally[33] "structural repeats larger than 30 or 40 residues are usually independent folding domains or modules that arrange as beads on a string", as we argued earlier based on the stability of Brownian motion[7]. The prominence of leucine in these repeats is not merely a packing effect, as in heptad zippers, as the two most common motifs are xLxL and LxxL[33]; it is not clear whether the prominence of leucine is a "hydrogen-bonding effect" or the result of "evolutionary infancy".

Leucine's prominence might be resolved in terms of functionality. Leucine terminates with a simple planar $CH_2$ group, which can occupy two orthogonal configurations, either containing the local chain axis, or normal to it. These two configurations are ideal for forming the simplest possible off/on (binary, and hence most efficient) bridging switch for H bonding to the water film associated with adjacent hydrophilic residues (always present because local hydroneutrality is required to have both stability and functionality). However, when the spacing of such switches is below a water-mediated correlation length at physiological temperatures, transition states become degenerate (within kT) in activation energy, and simple two-state transitions disappear. This effect has indeed been observed in designed LRR; by contrast, designed ankyrin repeats from the HEAT family (which have lower leucine densities) do exhibit two-state transitions[34]. Switches are the key to functionality, and it appears that the anomalously high frequency of leucine merely reflects its ideal capabilities as the simplest possible off/on (binary, and hence most efficient) bridging switch for H bonding to water films.

Here we will not undertake a proper statistical analysis of leucine heptads (a very large task), but it is worth remarking that a protein encoded by the human gene HEC (highly expressed in cancer) contains 642 amino acids and a long series of leucine heptad repeats (between aa 254 and 621, at its C-terminal region)[35]. We looked at two series (3 and 4 L's) of "pure L" heptad HEC repeats and found that the heptad $\langle\Psi 7\rangle$ values ranged from 0.153 to 0.107 – in other words, hydroneutral to strongly hydrophilic, suggesting that HEC leucine heptads are not primarily helical. This protein is now called HEC1; it is the subject of intense study, and hydroanalysis of the sequences of it and other proteins associated with it in the Ndc80 complex would be of interest[36].

Given the Zebende alphanumeric table for hydrophobicity, $-\gamma(aa)$, one can define a suitable average hydrophobiicty $\Psi(N,S) = \langle -\gamma(aa) \rangle$, where the average is taken either over N consecutive residues, or over part of, or an entire secondary element (such as a helix S). This average is similar to the one used by KD. A second inter-secondary, especially inter-helical quantity $\Phi$, is a measure of hydrophobic stiffness or flexibility,

$$\Phi(S) = \Sigma[(\gamma_{aa}(R(S)) - \gamma_{aa}(R(S+1)))^2 + (\gamma_{aa}(R(S)) - \gamma_{aa}(R(S-1)))^2]/2M \qquad (2)$$

Here R denotes an amino acid and $\gamma_{aa}(R)$ is its hydrophobicity. For applications to repeats, consecutive repeats are aligned in the standard matrix tableau dictated by the consensus set; the sum is over matched helices of maximum length M, so that both $\Psi$ and $\Phi$ are normalized. Note that consensus sites usually contribute 0 to $\Phi$, but that alignment of other sites may either increase or decrease $\Phi$, and several tests have shown that $\Phi$ is minimized by standard alignments in cases where the repeat proteins have crystallized. (However, $\Phi$ can also be utilized to align repeats that have not been crystallized, by minimizing the mismatch that $\Phi$ measures, thus overcoming some of the ambiguities associated with residue insertions or deletions.) Even in the context of adjacent (nearly parallel non-repeats, no consensus sites) α helices (such as merely H-bonded β strands), $\Phi$ could be useful. A general remark: overall there is a tendency towards hydroneutrality, but given proteins often separate into distinct hydroph(ob/il)ic regions, not only merely surface/volume, and these regions are easily identified with accurate values of $\Psi$.

5. **HEAT and ARM Repeat Proteins**

One of the most important differences between the present approach and many others is the emphasis here on functionality. Many bioinformatic searches span very large number of proteins in search of sequence-structure regularities, averaging over many functions. Here we are concerned mainly with identifying sequence-function relations. Of course, many sequence-structure-function relations are known, and we will utilize such relations wherever possible, but the sequence-function relations sought here are of a different nature. They are often not obvious, even when the structure is known, and as such can be called phantom relations. They arise because protein-protein interactions are usually mediated by water, and may be of longer range. Thus, even if it were possible to obtain pictures of proteins *in vitro*, one would still be uncertain of many aspects of protein-protein interactions. In practice one is limited in discussing sequence-structure-function relations to a few *ex vitro* crystallized examples. We shall see that these examples themselves are inherently biased towards certain kinds of functionality relative to others.

There are many families of proteins that one could choose to study, but the HEAT and ARM repeat proteins have many attractive features that are especially suitable for studying long-range water-mediated interactions. They have large SASA's that are easily separated into patches with distinct functionalities that are easily recognized with our phantom order parameters, much like liquid crystal domains. This becomes clearer if we compare these "open" repeats with slightly more compact solenoidal repeats.

HEAT and ARM repeats may utilize their large SASA's to perform multiple functions, a task that increases in importance from archea (< 1 repeat %) to metazoa (> 5 repeat %)[37]. These repeats exhibit functionality both mechanically and hydrophobically, in terms of the linear $\Psi$ and quadratic $\Phi$ order parameters defined above. The simplest mechanical aspects involve the formation of subhelices in each repeat. In the HEAT repeats there are typically 9 aa-specific consensus sites, and leucine occupies 5 of these (out of ~ 39 residues), and in the ARM repeats there are typically 8 aa-specific consensus sites, and leucine occupies 4 of these[38]. The predominance of leucine at consensus sites again can be explained in terms of H-bonded water mediated bridging switches.

Unlike leucine zippers and leucine-rich repeats, the helices of HEAT and ARM repeats split into sub-helices, labeled A and B for HEAT repeats. In the ARM repeats the A helices split into $H_{1,2}$ helices, and the B helices are relabeled $H_3$ helices. The A and B helices form an L-shape (like chopsticks), separated by a short (usually three residue) β turn[38]. The differences in functionality of HEAT and ARM repeats arise mainly from inter-repeat variations, and these appear differently in the spatial and hydrostructures. Thus these two families of repeats provide a severe test of whether it is the observed "real" (but *ex vitro*) spatial structure, or the "phantom" hydrostructure, that determines functionality.

## 6. Applications

Apparently Kyte and Doolittle's intentions were much the same as ours here: as they remark[30], the "simplicity and and graphic nature of hydrophobicity scales make them very useful tools for the evaluation of protein structures". Indeed, the KD scale has been widely cited. However, as noted before, it has many specific shortcomings, and the various inaccuracies have not inspired confidence in its quantitative reliability, or that of any alternative hydrophobicity scale. Here we rely almost entirely on the self-similar Zebende scale, for which we have already made many firm general supporting arguments, to which we can add many specific examples.

### A. Scaffolding HEAT Protein

As our first application, we consider the structure of PR65/A, a 590 residue subunit of protein phosphatase 2A, which contains 15 tandem repeats of ~ 39 residues each[38,39]. PR65/A is unusual because 90% of the aa's are helical (~ 73% is normal), and the A and B arms are symmetrical (18 aa's each). This unusually rigid and symmetrical structure is ideal for a scaffold supporting other subunits. The *ex vitro* marginal mechanical stability of the stressed protein is indicated by bends at the centers of both A and B arms, which also increase the volumes of their hydrophobic cores. Weakening of the helical structure is most pronounced in repeats 8-10, where 3-4 aa's are unwound. All repeats are nearly aligned, with the exception of the insertion of four hydrophilic aa's between repeats 8 and 9. The repeats form three blocks, with aberrant non-parallel packing between repeats 3 and 4, and 12 and 13, apparently caused by divergences from the consensus sequence; these divergences are described in terms of individual aa salt bridges, wedges, and other variations. In addition to leucines, a signature-motif sequence of the HEAT repeats of the PR65/A subunit and other HEAT motif proteins is the presence of conserved Asp and Arg (or occasionally Lys) residues; these are the three most hydrophilic aa's on the Zebende scale (not a coincidence, either for PR65/A, or the Zebende scale).

The $\Psi(A,B)$ and $\Phi(A,B)$ patterns of the A and B helical arms of PR65/A, shown in Fig. 4, are strikingly different, and reflect different functionalities. Generally speaking, hydro(phil/phob)ic interactions with water soften/harden helices; thus the left (N) half of PR65/A (Fig. 4(a)) exhibits hard comb-like $\Psi(B)$ arms and relatively soft $\Psi(A)$ arms, with a hydro hinge (both arms below hydroneutral at 0.155) at the central repeat $S = 8$. The A arms are associated with the convex outer surface, while the B arms belong to the concave inner surface, which functions as the scaffolding support[39] for catalytic and regulatory domains. The regulatory subunit B56γ1 (itself a 16-repeat) attaches[40] to repeats 2-7. The catalytic subunit attaches to the B arms of repeats 11-15, which are associated with large oscillations in $\Psi$. Thus the recently observed structure of the scaffold-regulatory-catalytic heterotrimer[40] partially reflects the inter-repeat spatial structure of the scaffold alone (the weakening of the helical structure in repeats 8-10 corresponds to the interface between the regulatory and catalytic units). However, at the same time the aberrant non-parallel packing between repeats 3 and 4, and 12 and 13, apparently caused by divergences from the consensus sequence, probably reflect only the marginal stability of isolated PR65/A. These "breaks" received detailed attention in discussions of the *ex vitro* structure, including many aa-specific packing interactions[38,39], but they appear to be irrelevant to the heterotrimer interfaces, which of course are tailored to *in vitro* functionality. The break at repeat 12 disappears in the heterotrimeric structure, in which the overall shape of PR65/A changes from a twisted hook to a more regular horseshoe[40].

Thus we turn to the (A,B) hydrofragility $\Phi_{A,B}$ patterns (Fig. 4(b)), which are quite different from the *ex vitro* packing patterns. Apart from softness near the N end, there is little structure between the A arms, but the B arms show distinct fragility peaks, at $S = 13, 4,$ and $12$ (in that order). One of the helical consensus sites in the B arm is 24, which is occupied by V in 11 out of the 15 repeats. This hydrophobic aa is missing from repeats 1 (N end), 4, 12 and 13, an essentially perfect correlation with the fragility peaks 13, 4, and 12. The marginal fragility of B arms of $S = 8$ and 9 is presumably correlated with the hydrohinge seen in $\Psi$ in Fig. 4(a). Overall it is clear that the $\Psi$(A,B) and $\Phi$(A,B) patterns of the helical arms of PR65/A, shown in Fig. 4, provide a better interpretation of the heterotrimer interfaces than the inter-repeat spatial patterns of isolated PR65/A.

The 3 aa turns at the vertex of the groove connecting the A and B arms are solvent-exposed, and it is useful to note that they form two distinct groups. Thus 10 of the turns are soft (strongly hydrophilic, $<\Psi 3> \sim 0.10$), while 5 are close to hydroneutral ($<\Psi 3> \sim 0.14$, repeats 3,4,8,10 and 14; note that four out of five of these turns are located near one of the three $\Phi_B$ peaks ). The latter may promote stability of the marginally stable scaffolding structure. Four oncogenic site mutations associated with lung and colon tumors have been discussed in terms of stability of the dry structure[39], but here one can propose an alternative *in vitro* interpretation. Two of the mutations occur in helical sites, and these two (P(65)→S and L(101)→P) involve substantial destabilizing decreases in hydrophobicity. The other two occur in the short turns, D(504)→G (repeat 13) and V(545)→A (repeat 14). The former stiffens the typically hydrophilic repeat 13 turn, while the latter softens the atypically nearly hydroneutral repeat 14 turn; both mutations regress $\Psi$ for these turns towards the mean value (0.118) for all 15 turns. All of these changes could easily favor more rapid production of the oncogenic protein, while disrupting its functionality. Another example[40] is E(64)→G (typically soft repeat 2), found in breast cancer, and in the same class as D(504)→G (repeat 13). (An example[40] from lung cancer is E(64)→D. The difference between D and E is very small, not only in hydrophobicity, but structurally they differ only by an extra backbone H, which is probably not significant.) These correlations are invisible in the context of the spatial structure of PR65/A alone[39], and are also not evident in the packing discussion of the heterotrimer interfaces[40]. Since these interfaces must be hydrated *in vivo*, it is not surprising that the specific details of the packing interactions in the heterotrimer crystal structure appear to have no functional content, whereas the functionality closely parallels both $\Psi$ and $\Phi$.

**B. Nuclear-Cytoplasmic Transport HEAT Proteins and Their Cargoes**

The general principle underlying cargo binding is matching of hydrophilic (basic) patches or segments. At first sight it might appear from an electrostatic model that such like-charged patches should repel (not attract), but in the presence of water they can share an intermediate water dielectric layer. Not only does such a layer shield the like charges, but it may even gain further energy by ordering into some kind of "ice-like" (four-fold coordinated oxygen) network. (In fact, inorganic interfacial networks (as in semiconductors) often exhibit such effects, and surface reconstructioned networks can be elaborate (the most famous example being the Si 7x7. Infrared studies of monolayer water films have shown such effects[41]). In practice philic-philic attraction is simply accepted as given. However, it is worth noting that to the extent that crystalline complexes are largely dehydrated, specific observed interfacial contacts may not be a good representation of the *in vitro* interrfacial structure. At the same time, the overall pattern should be correct: the cargo may shift slightly, but only slightly (by a few residues) *ex vitro*.

Transport of proteins from synthesis sites to functional sites is mediated by a multiplicity of repeat proteins, which typically load and unload their cargoes through large-scale conformational changes. The two most studied are importin α and β, and here we focus on importin β. When loaded, it has an S-like structure, whereas the empty structure is closed (globular)[42,43]. This is a large family, with at least 14 members in yeast and 22 in humans. The binding function of 19-repeat importin β (876 residues) involves an 9 aa hydrophilic loop between the A and B arms of repeat 8. This loop is strongly hydrophilic ($<\Psi 9> = 0.093$, deep in the hydrophilic aa tail, far below hydroneutrality at 0.155), but that is not all. While there is nothing special about $\Psi(A,B)$ at and near repeat 8 (Fig. 5(a))), there is a striking dip/peak (hard/soft) bifurcation in hydrofragility $\Phi(A,B)$ at repeat 8 (Fig. 5(b)). This is one of the largest A/B helical arm repeat asymmetries seen so far in our calculations.

The structure of 20-repeat transportin 1 (890 residues) is similar to that of importin β (24% sequence similarity); however, here there is a 60 aa hydrophilic loop ($<\Psi 60> = 0.118$, less hydrophilic than the importin β loop, but much longer) between the A and B arms of repeat 8. The structure of Trn1 has been determined, together with three cargoes[44], and many other cargoes of Trn 1 are known. These cargoes are an excellent proving ground for testing the importance of water interactions at protein interfaces.

First we take a closer look at the 60 aa hydrophilic loop; in Fig. 6 we see $<\Psi3>$, which exhibits a very interesting hydrostructure. Here the importin β 8 aa hydrophilic loop QDENDDDDD ($<\Psi9>$ = 0.093) reappears in Trn1 as EEEDDDDDE ($<\Psi9>$ = 0.091). It is supported by a chain of alternating [philic($<\Psi3>$ ~ 0.10)/neutral($<\Psi3>$ ~ 0.15)] links. Such an alternating structure is marginally stable, yet flexible enough to wrap around a variety of cargoes, while still utilizing the short 9 aa hydrophilic loop for actual binding. If the cargos are attached laterally to the transportin repeats, this alternating (not around hydroneutral, but neutral and deeply philic) chain could reinforce the binding provided by the loop by pinning the cargo with its philic links.

The next question is where the cargoes are attached. It appears that in the *ex vitro* crystal structures cargoes tend to collapse laterally onto hydrophilic residues of importin or transportin repeats. Thus it is suggested that the transportin binding sites for RanGTP lie along repeats 1-7, while there are two other binding sites (A, repeats 9-13 and B, 14-18) bind other cargoes, with repeats 14-20 being especially flexible. (This three-segment model has been discussed by many others.[45]. Some of the proposed hinges may be the result of buckling in the absence of water.) The unloaded Trn1 loop would collide with loaded repeat sites A and B if Trn1 were not deformed by loading, which suggests that the loop interacts strongly with cargoes bound to these sites, as confirmed by double mutagenesis experiments. Specifically cargoes TAP and JKTBP are bound to site A, while hnRNP is bound to both A and B[44].

Cargoes are attached to transporters through basic domains, and the transporters may fulfill a dual function, not only as nuclear import receptors, but also as cytoplasmic chaperones (suppressing aggregation) for exposed basic domains[45,46]. Cargo basic domains also have characteristic hydrostructures. By far the most studied cargo is the GTP-binding protein Ran, which in turn regulates transporters' interaction (loading and unloading in the nucleus and cytoplasm) with cargo. *Ex vitro* structural studies[47,48] suggest that Ran is bound to the (convex inner) B arms of repeats 1-7 of importin β, as well as its acidic loops, but the Ran cargo may have shifted relative to the repeats of importin β to *in vivo* upon dehydration and even crystallization. Thus comparison of the hydrostructure of importin β relative to that described by *ex vitro* structural studies is of great interest.

**C. Cargo Hydrosignatures**

The hydrosignature of Ran is complex, as befits such an active and important regulatory protein. Thus we begin our discussion of cargo hydrosignatures with simpler cases, to reassure readers who may be uncertain of the reliability of hydroanalysis for establishing protein binding areas. Multiple ribosomal proteins play modest (but essential!) supporting roles by complexing with RNA during protein synthesis, but they must avoid aggregation. Figs. 7-9 display the hydrosignatures of three ribosomal proteins[46], S7, L23a, and L18a. The residue patch 98-120 [102-120] of S7 has a formal charge[46] of +12 [+10], but we prefer to say that its $<\Psi 22>$ [$<\Psi 19>$] is 0.103 [0.094], both because such a large charge accumulation is unlikely to be accurately measured simply by using pI values, and because the hydroscale is smooth, seamless, dimensionless, and simply more accurate for making comparisons to other proteins. Note the deep philic valley associated with this patch

Our next example, L23a, does not exhibit an obvious basic patch, and is nearly hydroneutral ($<\Psi 176> = 0.147$), but it exhibits a striking internal hydrostructure (Fig. 8), which is particularly well-suited to interaction with a repeat protein like importin β. Between two hydrophobic peaks there is a ~ 35-residue structure that oscillates between hydroneutral ($<\Psi 3> \sim 0.15$) and philic ($<\Psi 3> \sim 0.09$). The six philic dips, placed about five residues apart, could easily attach to basic residues on importin β repeats, and they resemble the basic chain of transportin discussed above. Fragmentation experiments[49] established that the functionally active basic domain is between residues 32 and 74, essentially a perfect match to the camel-back structure between the two strong hydrophobic peaks in Fig. 8.

L18a is the most marginal example (Fig. 9). At first sight no basic patch is obvious, and none was identified, but having seen the 35-residue camel-back structure of L23a, we can now recognize in L18a a very wide and less dramatic 100-residue triple camel-back structure, with two sets of philic ($<\Psi 3> \sim 0.09$) dips. Here the neutral peaks are replaced by phobic peaks, but the functionality (in terms of avoiding aggregation by complexing with importin β) appears to be much the same. This tells us that it is the philic dips that are bound to importin β, and the peaks are less important.

Now that we are familiar with some of the many possible large-scale importin β binding patterns, we are prepared to look at *ex vitro* Ran-importin β binding[47,48]. The B1-B3 (and B6-B7) repeat arms of importin β1 wrap around residues 64-113 (and 141-173) of Ran. Also the disordered Ran C-terminal philic helix (DEDDD) presumably binds to the importin β1 philic patch DENDDDDD. Importin β2 contains a small philic patch DVEEDE similar to that of importin β1, but it also exhibits a much larger 19-residue philic patch DEDGIEEEDDDDDEIDDDD associated with loop 7, which produces large structural differences between the two importin β-Ran complexes, especially *ex vitro*. As the hydration shifts of Ran relative to importin β1 are expected to be much smaller than those of importin β2, we focus our hydrostructure discussion on the former; for discussion the Ran hydrostructure has been split into three parts (Fig. 10 (a)-(c)).

The philic nature of the Ran C-terminal helix (DEDDD) is obvious in Fig. 10(c). The 64-113 Ran section bound to importin β1 repeats B1-B3 is included in Fig. 10(a); it has $<\Psi 50> = 0.156$ (hydroneutral). The most interesting region, including binding to the B7, B8 importin β arms, is shown in Fig. 10(b); the reader will recognize immediately the camel-back structure between peaks P1 (136) and P2 (188). Between these two peaks there are six "tilted" philic minima, which decrease in depth monotonically from $<\Psi 3> = 0.072$ at 141, to $<\Psi 3> = 0.112$ at 186. The first five of these minima match the described *ex vitro* structure well, and it is not surprising that the sixth and weakest minimum is obscured by the disorder associated with the A8-B8 "stalk"[48]. The "tilted" structure of the camel back valley strongly suggests an elegant function for this broad valley. In binding, the usual rule (expected from enthalpy-entropy compensation, which is well-obeyed in protein solute-iceberg experiments[50]) is that the highest minimum is accessed first (it has the largest entropy), followed by a cascade to lower minima. Here this would mean that binding would begin with the entanglement of the basic importin β and Ran loops, and then would be followed by successive "domino" contacts across the tilted Ran camel-back philic minima. Unbinding would involve disentangling the exposed loops, followed by peeling off successive "domino" contacts.

Because the central section of Ran (the "basic patch") yields such a pretty tilted camel's back hydrostructure, it is an ideal platform for comparing the Zebende self-similar scale[27] to other hydrophobicity scales. In Fig. 11(a) we reproduce our benchmark Fig. 10(b), followed in Fig. 11(b) by a plot using the "amalgam" scale[30], and in Fig. 11(c) the buried scale[24]. At first glance the "amalgam" scale seems to do a better job of reproducing the benchmark, but this is only because the eye tends to follow the hydrophobic peaks. If one looks more closely at the hydrophilic minima (which are functionally more important), then it is clear that the buried scale is better than the "amalgam" scale. However, compared to the Zebende benchmark, neither of the other scales accurately reproduces the perfectly monotonic (both phobic and philic) tilted camel's back hydrostructure of Ran.

Because hydrostructure is defined by functionality, in general it should have little or nothing to do with *ex vitro* stability. The repeat proteins are in this respect exceptional, because their structures are stabilized by water-mediated inter-repeat interactions. This is not the case, however, for Ran, which is not a repeat protein and instead resembles more closely a globular protein. By examining the beta strand, helix and turn structure of the central basic section of Ran, we found that, as expected, there is little or no correlation between *ex vitro* stability and hydrostructure. Since the latter provides an elegant picture of functionality, we conclude that it is easier to understand the functionality of proteins by studying their phantom hydrostructures; indeed, many attempts that been made to explain functionality from spatial structure have failed, leaving grave (and well justified!) doubts as to whether functionality could be explained by CFF.

The final examples come from the heterogeneous nuclear ribonucleoproteins hnRNP A1 and D. These cargo proteins are glycine- and alanine-rich (30% or more), which makes them very flexible; they are transported by importin β and by transportin[51], respectively. These two cargoes show different nuclear localization signals (NLSs); both terminate with hydrophobic residues PY, but the NLS of A1 is 30-40 aa's long, whereas the NLS of D is 19 aa's long[51]. Is this difference due to different transporters [the shorter basic loop in importin β and the longer one in transportin], or can it be recognized already in the respective cargo hydrostructures? The answer is shown in Fig. 12. From 316 to 340-1(PY), the hydrostructures are quite similar, but A1 has 313-314 = AD with $<\Psi2> =0.100$, while D has 313-4 = YG with $<\Psi2> =0.189$. From the figure this difference is obvious, although it is slightly obscured by using a 3- (instead of 2-) point average, but without the convenience and reliability of the Zebende absolute hydroscale, it was easily missed[51].

## 7. Conclusions

Our primary thesis – that protein functionality is determined by water-protein interactions – has been implemented by utilizing the Zebende absolute hydroscale (analogous to the absolute dielectric ionicity scale) to analyze the properties of repeat proteins and their interactions with other proteins – notably the transport repeat proteins interacting with a wide variety of cargoes. The results appear to be excellent. The techniques used involve extremely simple calculations that were carried out using EXCEL on a PC. Each example took only a few minutes of (human) calculation (much more time was spent exploring the literature, but important papers were surely still left uncited). The reader may wonder how many cases were explored, and if the excellence of the results is not due to omission of weak examples. In fact, the examples shown are not just typical – they are all the cases studied (there were no weak examples)! In other words, interesting results were obtained for every case studied. There are many more cases to be studied in the future – the ARM repeat family, closely related to the HEAT family, and the many subtle differences between the many members of each family.

It appears that evolutionary connections between hydrostructure and functionality should be accessible, and there is no reason the method (which worked just as well for the globular glycine-alanine-rich repeats as for simpler globular cargoes) should not be quite general. In principle, for example, hydrofragility should provide a powerful additional tool for aligning helices in globular proteins; if it is focused on all the representatives of a given fold, then that would provide a simple and presumably powerful step forward towards solving the folding problem. The possible applications of the method appear to be limited only by the scope of protein data, and as systematic patterns emerge, the technique could easily acquire substantial predictive power.

My interest in this subject was originally stimulated by the pioneering mutagenesis studies of transition states by Prof. A. R. Fersht, which in turn led me to examine the logical relations shown in Fig. 1. The proximate source of my interest was elegant mutagenesis transition-state studies of ankyrin repeats[52], which exhibit large-scale collective behavior, suggestive of an order-parameter approach. This in turn led me to the Pintar-Zebende scales; the dimensionless nature of the latter is highly appealing, especially in the light of Boolchand's discovery of the reversibility window in network glasses[8,9]. Even so, I have been very pleasantly surprised by the absence of failures with this method. Enlightening advice on the current state of quantum theories of H bonding from Prof. A. Karpfen are gratefully acknowledged.

## References

1. L. Pauling, *The Nature of the Chemical Bond* (Cornell, 1960). p. 93.

2. M. L. Tiago, S. Ismail-Beigi and S. G. Louie, J. Chem. Phys. **122**, 094311 (2005).

3. J. C. Phillips, Rev. Mod. Phys. **42**, 317 (1970).

4. J. C. Phillips, *Bonds and Bands in Semiconductors* (Academic Press, New York, 1973).

5. F. Seitz, *The Modern Theory of Solids* (McGraw Hill, New York, 1940).

6. H. Abu-Farsakh and A. Qteish, Phys. Rev. B **75**, 085201 (2007).

7. J. C. Phillips, J. Phys. Cond. Mat. **16,** S5065-S5072 (2004).

8. S. Chakravarty, D. G. Georgiev, P. Boolchand, and M. Micoulaut, J. Phys. Cond. Mat. **17**, L1 (2005).

9. D. Novita, P. Boolchand, M. Malki and M. Micoulaut, Phys. Rev. Lett. **98**, 195501 (2007).

10. J. R. Forman and J. Clarke, Curr. Opin. Struc. Biol. **17**, 58 (2007).

11. N. V. Dokholyan, Gene **347**, 199 (2005).

12. Y. Wang, J. Wells, D. G. Georgiev, P. Boolchand, K. Jackson, and M. Micoulaut, Phys. Rev. Lett. **87**, 185503 (2001).

13. G. Lucovsky and J. C. Phillips, J. Phys. Cond. Mat. **19**, 455218 (2007).

14. J. C. Phillips, J. Phys. Cond. Mat. **19**, 455213 (2007).

15. O. D. Monera, C. M. Kay and R. S. Hodges, Prot. Sci. **3**, 1984 (1994).

16. A. Karpfen, Adv. Chem. Phys. **123**, 469 (2002).

17. K. Y Kim, H. J. Lee, A. Karpfen, J. Park, C. J. Yoon and Y. S. Choi, Phys. Chem. Chem. Phys. **3**, 1973 (2001).

18. Y. F. Chen, R. Viswanathan, and J. J. Dannenberg, J. Phys.Chem. B **111**, 8329 (2007).

19. R. P. Sheridan, R. H. Lee, N. Peters and L. C. Allen, Biopoly. **18**, 2451 (1979).

20. L. Ismer, J. Ireta, S Boeck and J. Neugebauer, Phys. Rev. E **71**, 031911 (2005).

21. R. Wieczorek and J. J. Dannenberg, J. Phys.Chem. B **112**, 1320 (2008).

22. S. Scheiner, J. Phys. Chem. B **111**, 11312 (2007).

23. R. L. Baldwin, J. Mol. Biol. **371**, 283 (2007).

24. A. Pintar, O. Carugo and S. Pongor, Trends Biochem. Sci. **28**, 593 (2003); Biophys. J. **84**, 2553 (2003).

25. Z. Yuan and Z.-X. Wang, Proteins- Struc. Func. Bioinfo. **70**, 509 (2008).

26. R. Eudes, K. Le Tuan, J. Delettre, J._P. Mornon and I. Callebaut, BMC Struc. Biol. **7**, 2 (2007).

27. M. A. Moret and G. F. Zebende Phys. Rev. E **75**, 011920 (2007).

28. A. Soyer, J. Chomilier, J.-P Mornon, R. Jullien, and J. F. Sadoc, Phys. Rev. Lett. **85**, 3532 (2000).

29. N. Pittet, J. Phys. A **32**, 4611 (1999).

30. J. Kyte and R. F. Doolittle, J. Mol. Biol. 157, 105 (1982).

31. F. H. C. Crick, Acta Cryst. **6**, 689 (1953).

32. J. Liu, Q. Zheng, Y. Deng, C.-S. Cheng, N. R. Kallenbach and M. Lu, Proc. Nat. Acad. Sci. (USA) **103**, 15457 (2006).

33. B. Kobe, Nature Struct. Biol. **3**, 977 (1996).

34. H. K Binz, A. Kohl, A. Plueckthun and M. G.Gruetter, Proteins – Struct. Func. Bioinfo. **65**, 280 (2006).

35. Y. M. Chen, D. J. Riley, P. L. Chen and W.-H. Lee, Mol. Cell. Biol. **17**, 6049 (1997).

36. J. G. DeLuca, W. E. Gall, C. Ciferri, D. Cimini, A. Musacchio and E. D. Salmon, Cell **127**, 969 (2006).

37. M. A. Andrade, C. Perez-Iratxeta, C. P. Ponting, J. Struct. Biol. **134**, 117 (2001).

38. B. Kobe, T. Gleichmann, J. Horne, I. G. Jennings, P. D. Scotney and T. Teh, Struc. **7**, R91 (1999).

39. M. R. Groves, N. Hanlon, P. Turowski, B. A. Hemmings and D. Barford, Cell **96**, 99 (1999).

40. U. S. Cho and W. Xu, Nature **445**, 53 (2007); Y. M. Chook and G. Blobel, Curr. Opin. Struc. Biol. **11**, 703 (2001).

41. M. C. Gurau, G. Kim, S. M. Lim, F. Albertorio, H. C. Fleisher and P. S. Cremer, Chem. Phys. Chem. **4**, 1231 (2003).

42. A. Cook, E. Fernandez, D. Lindner, J. Ebert, G. Schlenstedt and E. Conti, Mol. Cell **18**, 355 (2005).

43. E. Conti, C. W. Muller and M. Stewart, Curr. Opin. Struct. Biol. **16**, 237 (2006).

44. T. Imasaki, T. Shimizu, H. Hashimoto, Y. Hidaka, S. Kose, N. Imamoto, M. Yamada and M. Sato, Mol. Cell **28**, 57 (2007).

45. A. E. Cansizoglu and Y. M. Chook, Struc. **15**, 1431 (2007).

46. S. Jakel, J. M. Mingot, P. Schwarzmaier, E. Hartmann and P. Gorlich, Embo **21**, 377 (2002).

47. Y. M. Chook and G. Blobel, Nature **399**, 230 (1999).

48. I. R. Vetter, A. Arndt, U. Kutay, D. Gorlich and A. Wittinghofer, Cell **97**, 635 (1999).

49. S. Jakel and P. Gorlich, Embo **17**, 4491 (1998).

50. J. Kang and A. S. Warren, Mol. Imm. **45**, 304 (2008); J. C. Phillips, J. Chem. Phys. **81**, 478 (1984) ; T. Imai and F. Hirata, J. Chem. Phys. **122**, 094509 (2005); Y. Koga , K. Nishikawa and P. Westh, J. Phys. Chem. A **108**, 3873 (2004).

51. T. Imasaki, T. Shimizu, H. Hashimoto, Y. Hidaka, S. Kose, N. Imamoto, M. Yamada and M. Sato, Mol. Cell **28**, 57 (2007).

52. N. D. Werbeck and L. S. Itzhaki, Proc. Nat. Acad. Sci. (USA) **104**, 7863 (2007).

**Figure Captions**

Fig. 1. A sketch of the titled topics and their inter-relations. Note that it is our position that even if a way could be found to calculate protein stability reliably using classical force fields, that same CFF would not yield a description of protein functionality.

Fig.2. A sketch illustrating the importance of reversibility, and why stability and functionality are probably not accessible in proteins with a common formalism. The edges of the inverted parabolic spinodal separating stable fully folded proteins from denatured proteins define two phase transitions, which have been identified in studies of the reversibility window in network glasses[8,9]. If the intermediate region is narrow enough, the two transitions appear to merge, and the protein is said to undergo a two-stage transition from folded to unfolded. With a broader intermediate region, the transition is called three (or more)-stage. In any case, models that describe the stability of the folded state are unsuited to describing functionality in the intermediate region, as elaborated further in the text.

Fig. 3. Three water-only alphanumeric hydrophobicity scales, designed to represent hydrophobic effects without expanding in terms of clusters and their interaction energies. Amino acids are coded by letters, and the three scales have been adjusted to span similar ranges with similar averages (hydroneutrality) near 0.155. The entries in order correspond to (1) "amalgam"[30] (2) mean depth[24] and (3) self-similar SASA[27]. While scales (1,2) have been tested against scale (3), the latter is regarded as best and is the only scale used here, apart from Fig. 11.

Fig. 4(a,b). $\Psi$ and $\Phi$ diagrams for the scaffold protein PR65/A (in color on line): A (B) helical arms are in blue diamonds (red squares). Note that $\Phi$(B) displays three peaks. The 2-7 peak binds a regulatory subunit, a catalytic subunit is bound to 11-15, and the central $\Phi$ peak is associated with the interface between these two subunits.

Fig. 5(a,b). $\Psi$ and $\Phi$ diagrams for importin β1 (in color on line): A (B) helical arms are in blue diamonds (red squares).

Fig. 6. The hydrostructure of the 60 aa loop of transportin.

Fig. 7. The hydrostructure of the ribosomal protein rpS7.

Fig. 8. The hydrostructure of the ribosomal protein rpL23a.

Fig. 9. The hydrostructure of the ribosomal protein rpL18a.

Fig. 10(a)-(c). The hydrostructure of Ran is presented in three parts. To make the connections easier, the two largest hydrophobic peaks are labeled P1 and P2.

Fig. 11. The hydrostructure of the Ran camel back section usiong (a) the dimensionless self-similar Zebende scale[27] (b) the "amalgam" energy scale[30] and (c) the buried length scale[24].

Fig. 12. The hydrostructure of the heterogeneous nuclear ribonucleoproteins hnRNP A1 and D (in color on line).

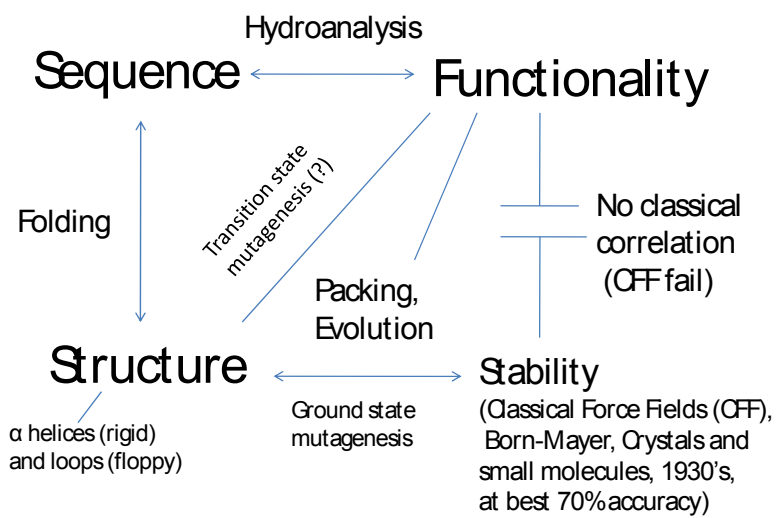

Fig. 1.

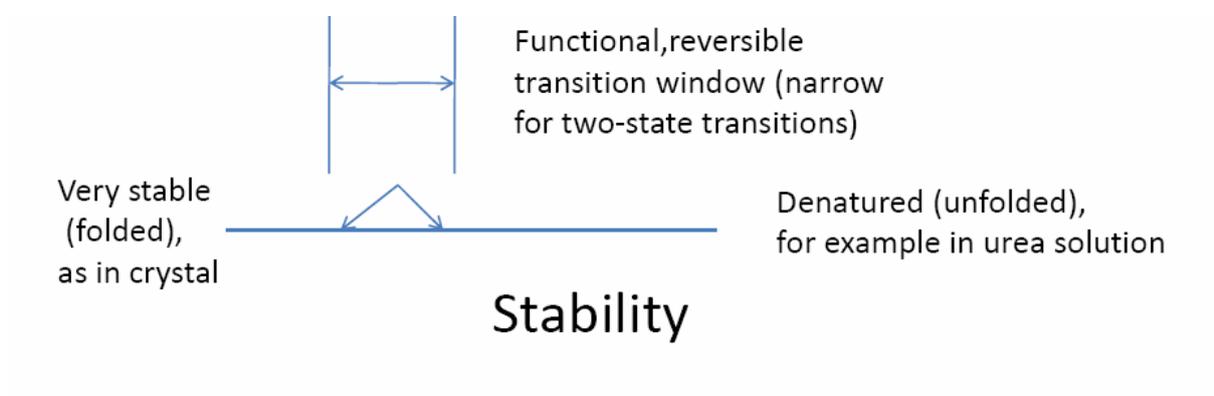

Fig. 2

| A | 0.200 | C | 0.214 | D | 0.096 | E | 0.096 | F | 0.220 |
|---|---|---|---|---|---|---|---|---|---|
|   | 0.156 |   | 0.179 |   | 0.077 |   | 0.068 |   | 0.219 |
|   | 0.157 |   | 0.246 |   | 0.087 |   | 0.094 |   | 0.218 |
| G | 0.157 | H | 0.101 | I | 0.253 | K | 0.088 | L | 0.240 |
|   | 0.102 |   | 0.122 |   | 0.246 |   | 0.064 |   | 0.217 |
|   | 0.156 |   | 0.152 |   | 0.222 |   | 0.069 |   | 0.197 |
| M | 0.202 | N | 0.096 | P | 0.133 | Q | 0.096 | R | 0.076 |
|   | 0.196 |   | 0.087 |   | 0.097 |   | 0.083 |   | 0.087 |
|   | 0.221 |   | 0.113 |   | 0.121 |   | 0.105 |   | 0.078 |
| S | 0.149 | T | 0.157 | V | 0.248 | W | 0.147 | Y | 0.139 |
|   | 0.100 |   | 0.117 |   | 0.225 |   | 0.197 |   | 0.169 |
|   | 0.100 |   | 0.135 |   | 0.238 |   | 0.174 |   | 0.222 |

Fig. 3.

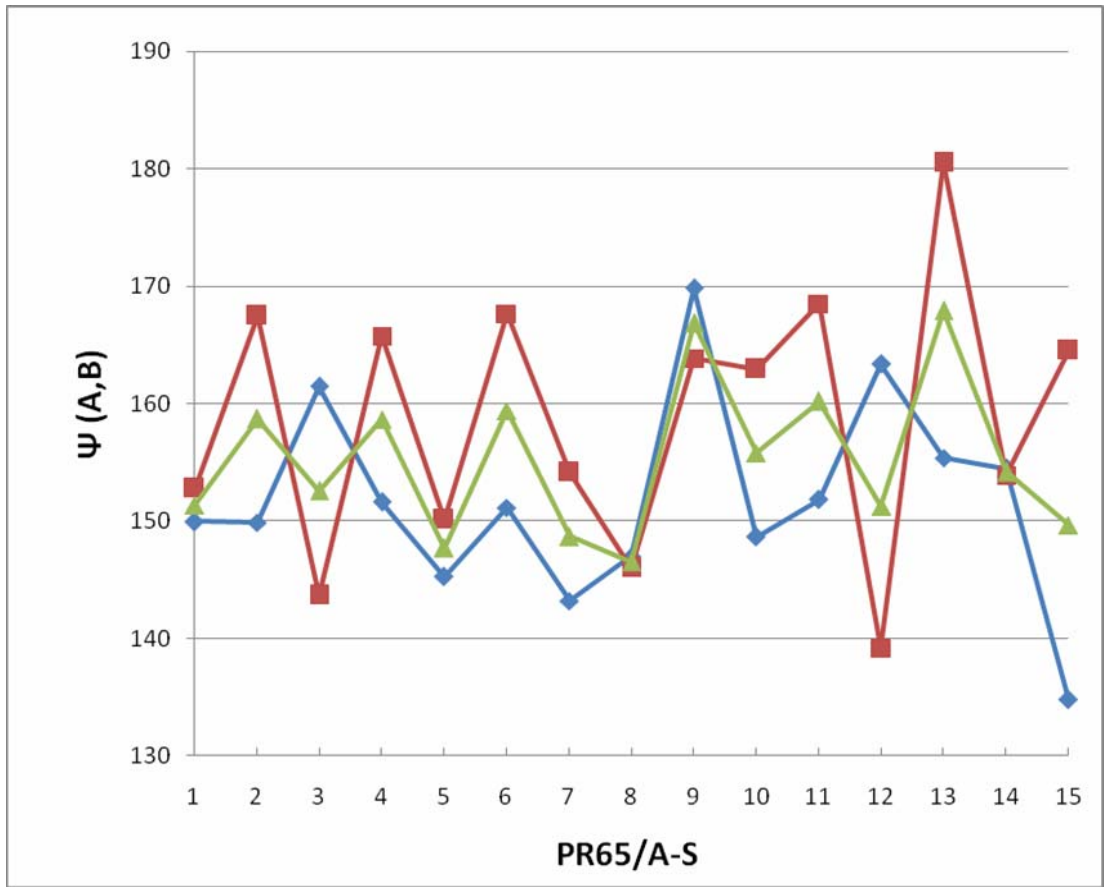

Fig. 4(a).

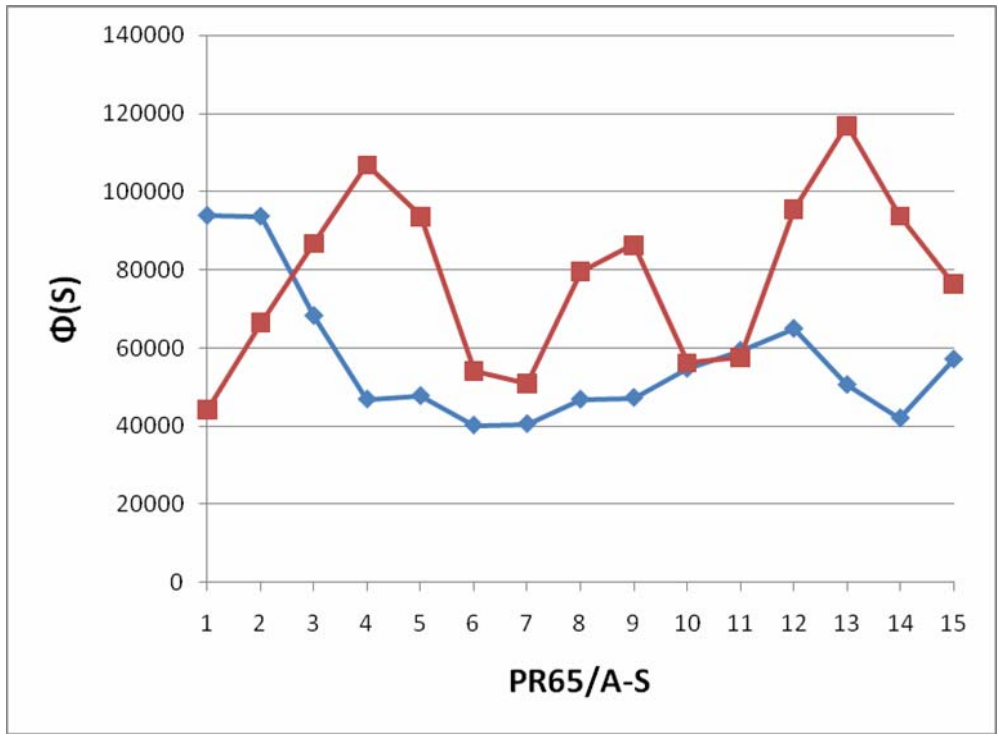

Fig. 4(b).

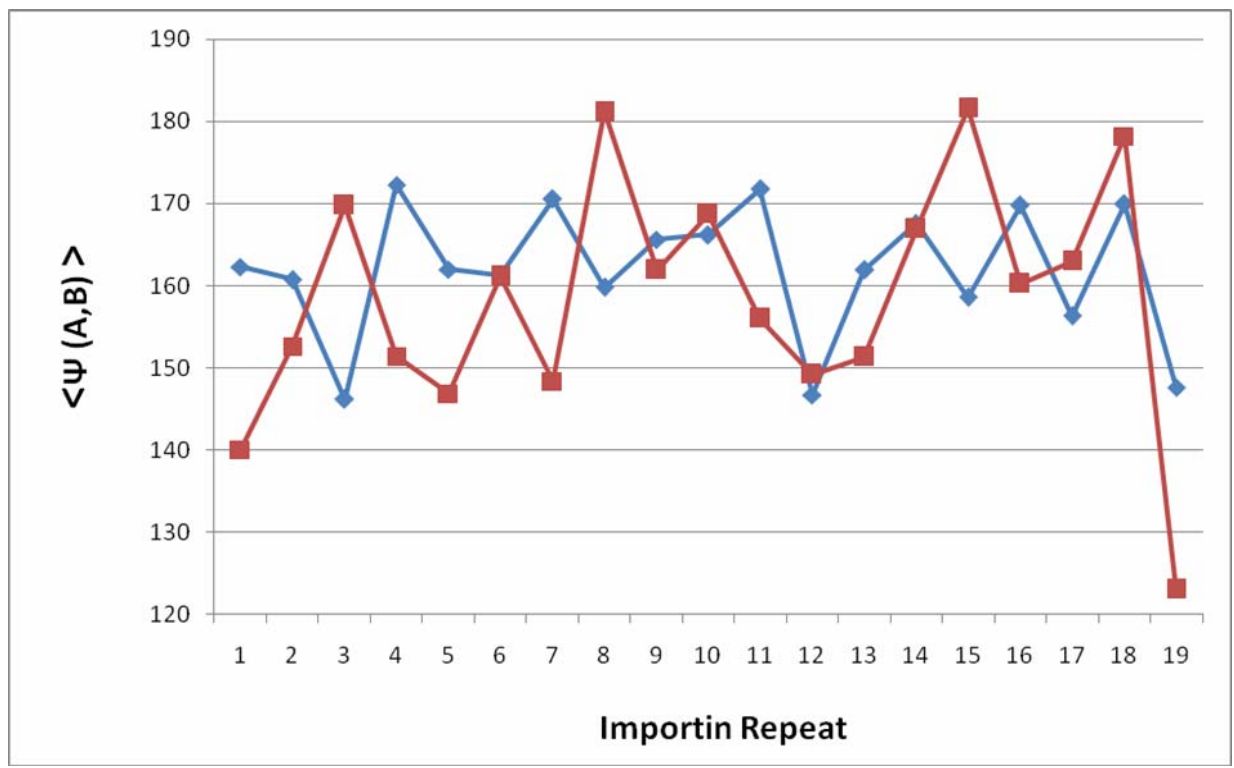

Fig.5(a).

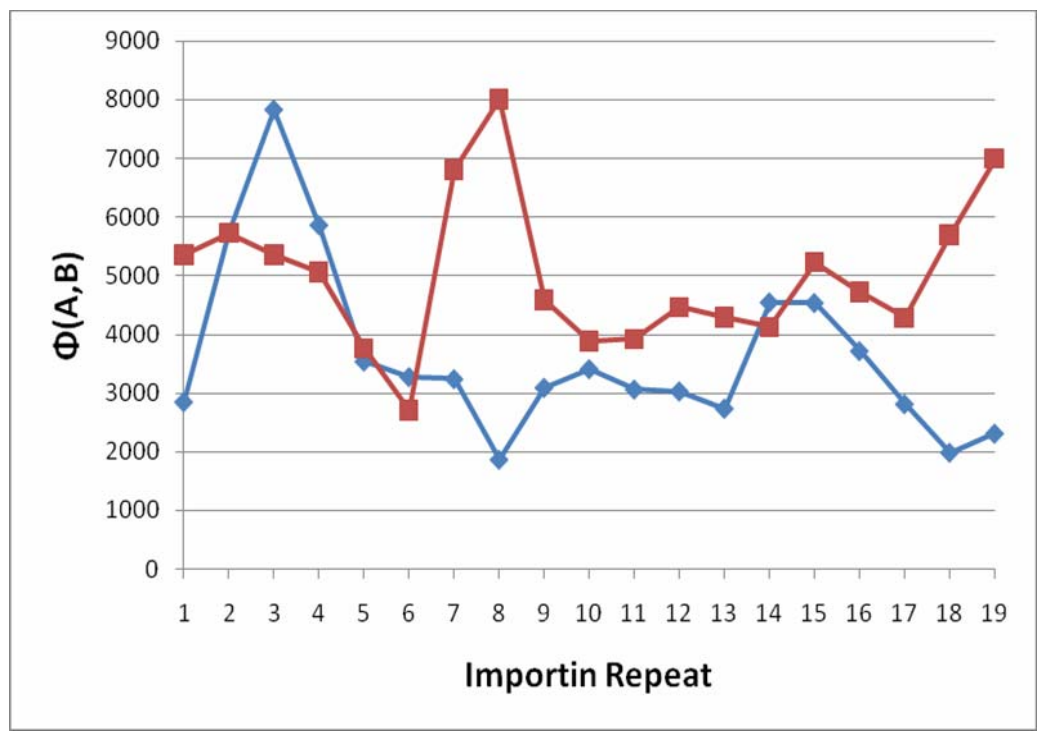

Fig. 5(b).

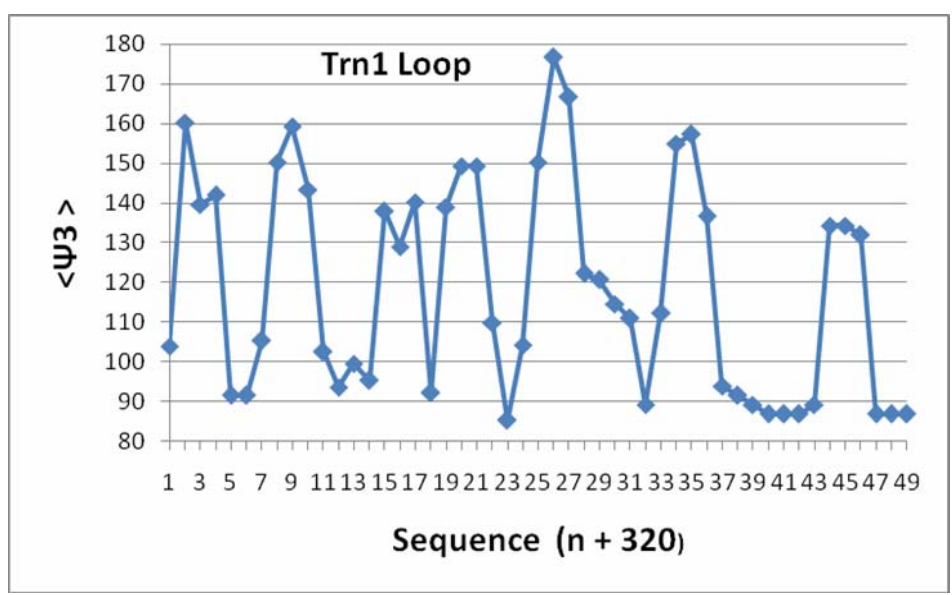

Fig.6.

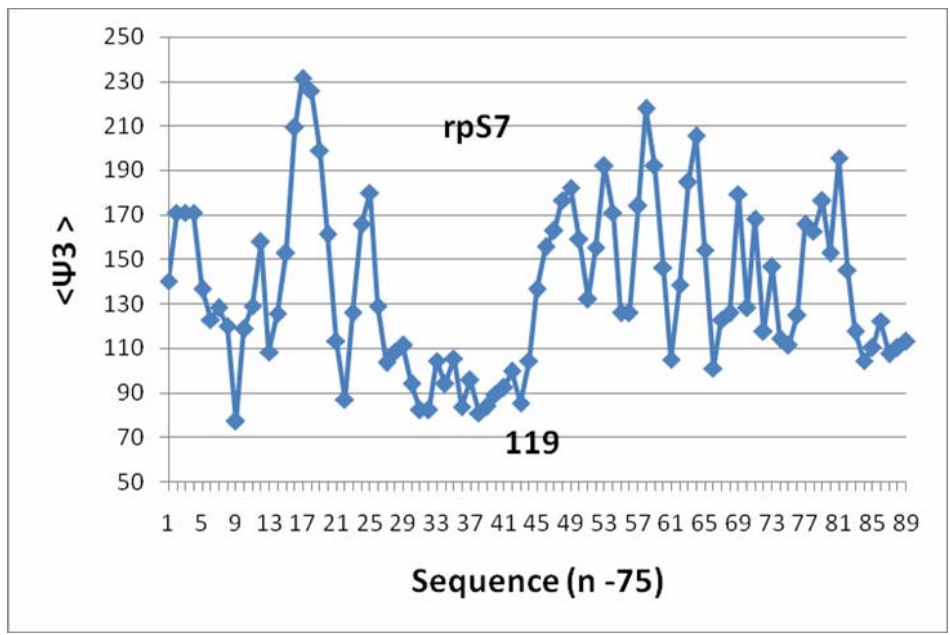

Fig. 7.

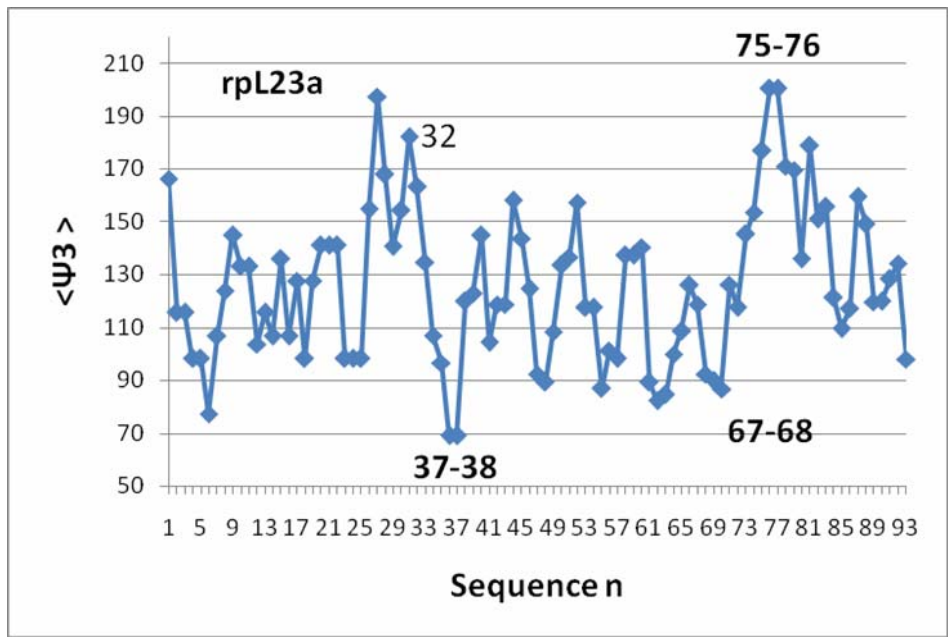

Fig. 8.

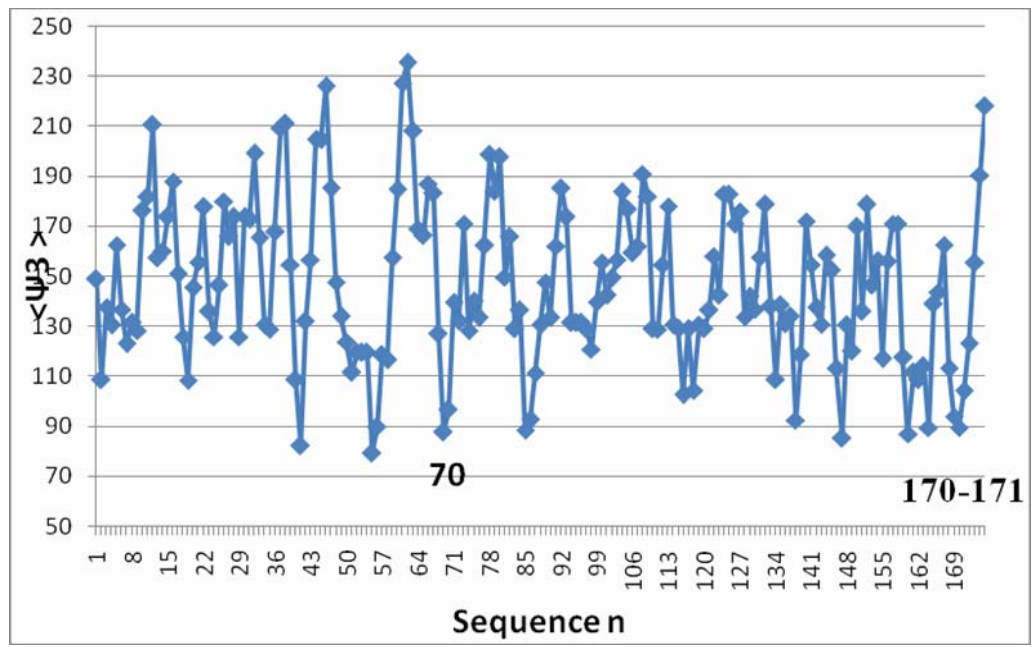

Fig. 9.

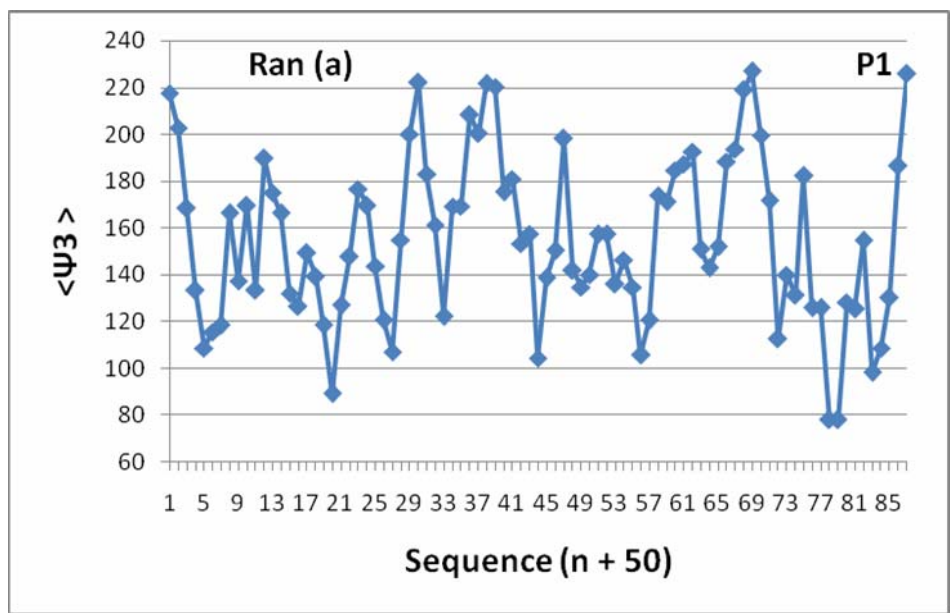

Fig. 10(a).

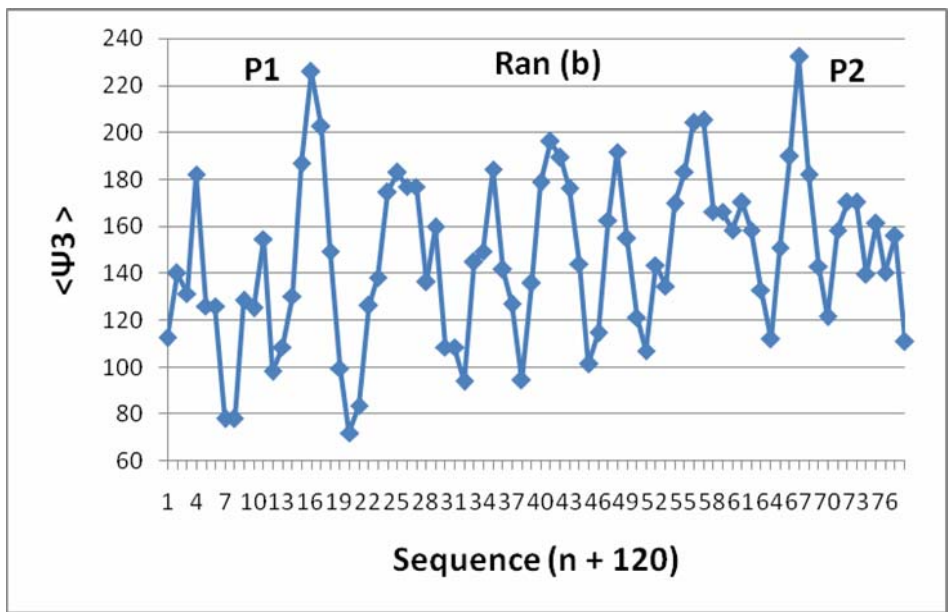

Fig. 10 (b)

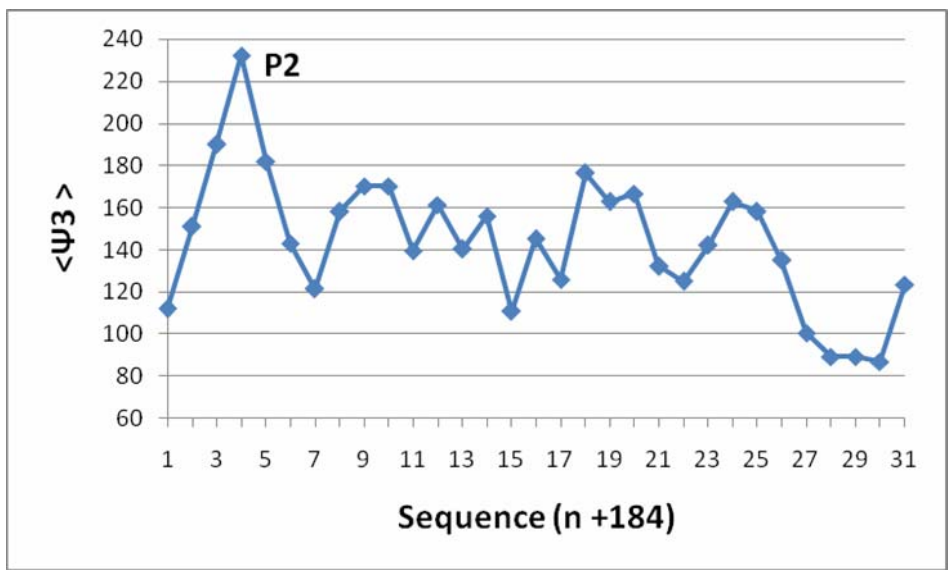

Fig. 10(c).

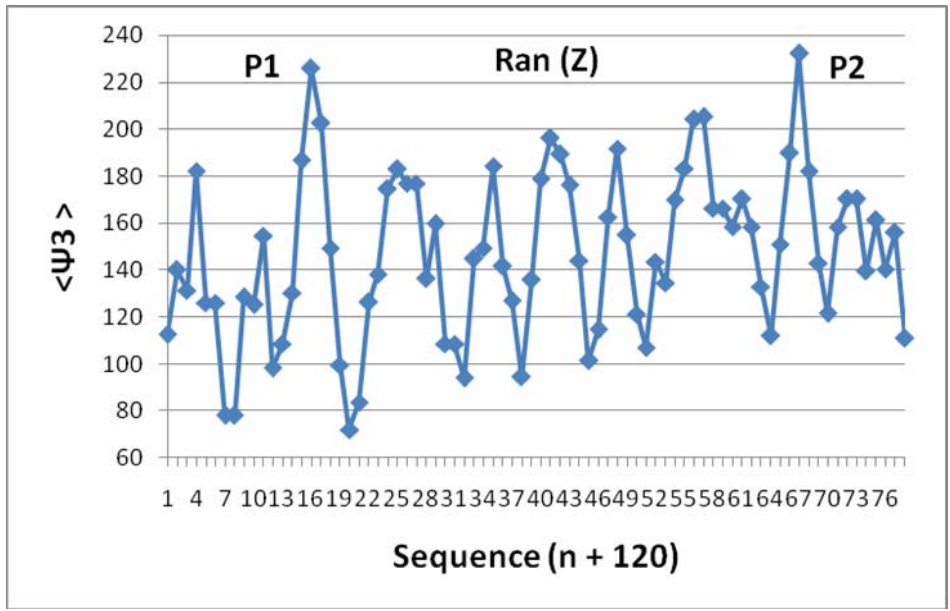

Fig. 11(a).

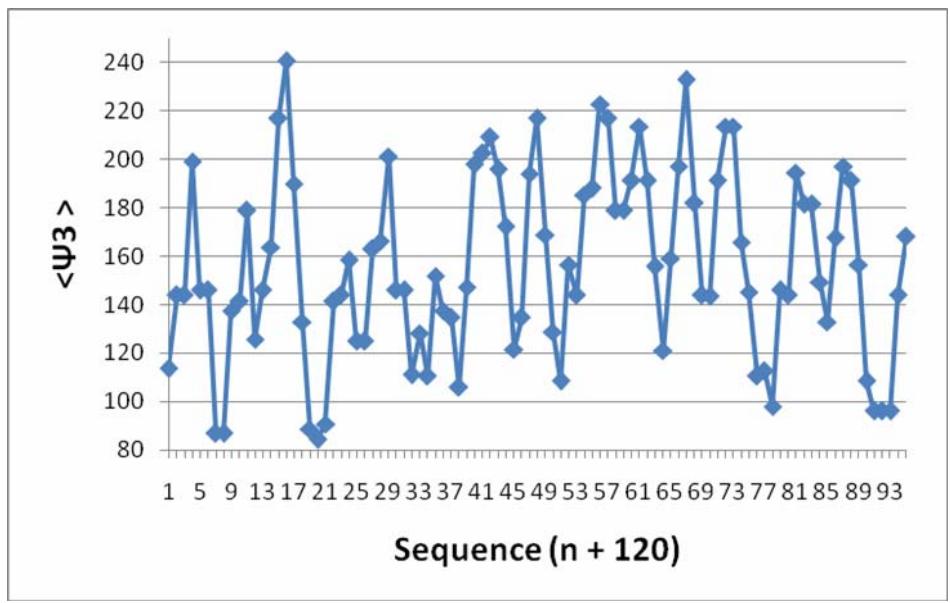

Fig. 11(b).

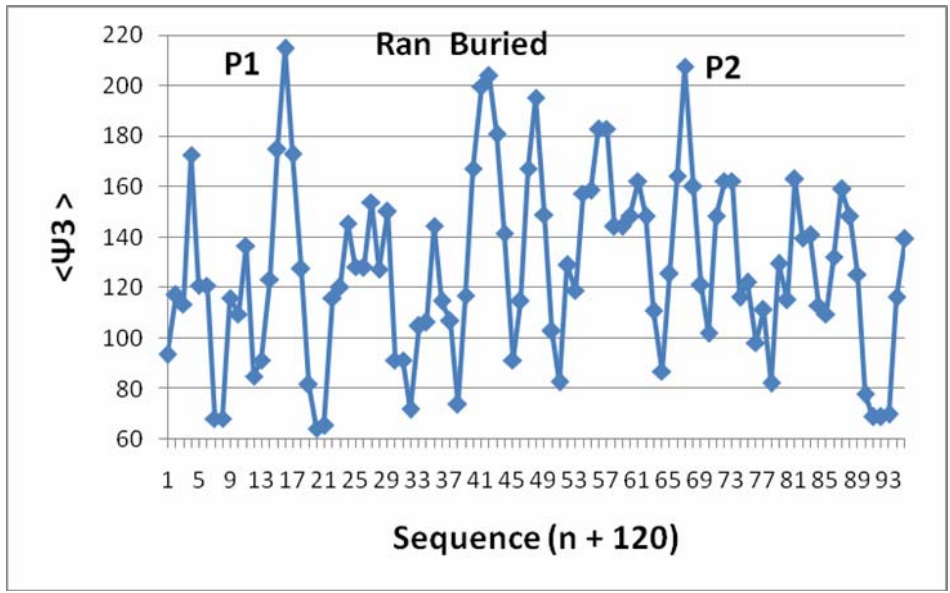

Fig. 11(c).

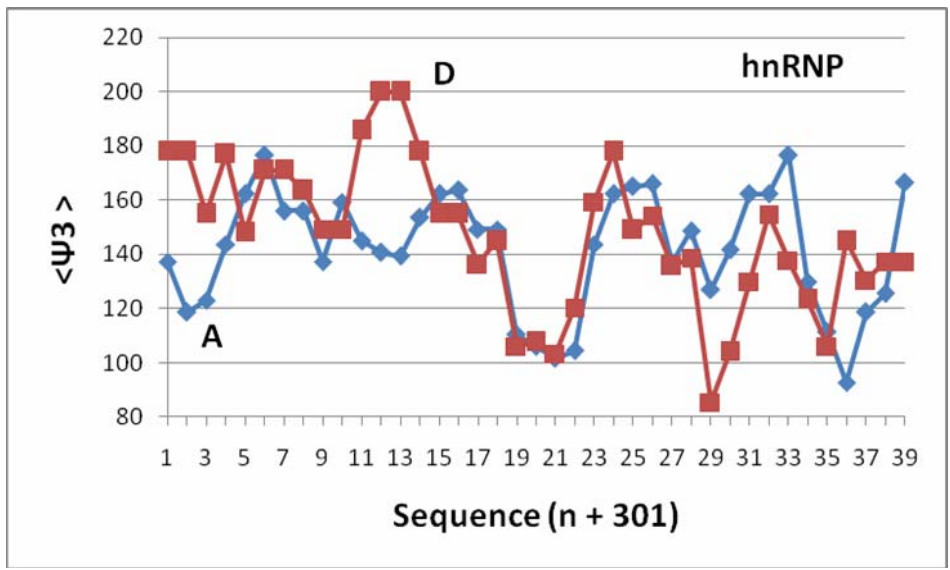

Fig. 12.